\documentstyle[prl,twocolumn,aps]{revtex}
\RequirePackage{epsfig}
\def\Journal#1#2#3#4{{#1} {\bf #2}, #3 (#4)}
\def\PR{Phys. Rev.}
\def\PRL{Phys. Rev. Lett.}
\def\PRA{Phys. Rev. A}

\def\RMP{Rev. Mod. Phys.}
\def\JMP{J. Math. Phys.}
\def\JPUSSR{J. Phys. (U.S.S.R.)}
\def\NC{Nuovo Cimento}
\def\ZETF{Zh. Eksp. Teor. Fiz.}
\def\JETP{Sov. Phys. JETP}
\def\APNY{Ann. Phys. (N.Y.)}

\begin{document}
\draft
\title{Bose-Fermi variational theory of the BEC-Tonks crossover}
\author{M. D. Girardeau\thanks{Email: girardeau@optics.arizona.edu} and
E. M. Wright\thanks{Email: Ewan.Wright@optics.arizona.edu}}
\address{Optical Sciences Center and Department of Physics,
University of Arizona, Tucson, AZ 85721}
\date{\today}
\maketitle
\begin{abstract}
A number-conserving hybrid Bose-Fermi variational theory is
developed and applied to investigation of the BEC-Tonks gas
crossover in toroidal and long cylindrical traps of high aspect
ratio, where strong many-body correlations and condensate
depletion occur.
\end{abstract}
The many-body ground state of a trapped atomic vapor Bose-Einstein
condensate (BEC) is described in first approximation by
Gross-Pitaevskii (GP) theory \cite{Gross,Pitaevskii} and in the
next approximation by Bogoliubov theory \cite{Bog} or
time-dependent GP theory \cite{DGPS}. These fail if the condensate
is appreciably depleted, as is the case near the BEC condensation
temperature, or even at $T=0$ for sufficiently thin wave guides,
low densities, and large scattering lengths
\cite{Olshanii,Petrov}. An extreme limiting case of this, the
``Tonks gas'' where transverse excitations are frozen, the
dynamics reduces to one-dimensional (1D) motion, BEC disappears,
and the occupation $N_0$ of the lowest orbital behaves like $N^p$
with $p<1$, can be treated exactly by the Fermi-Bose mapping
method \cite{map,GW3}. The behavior of the BEC-Tonks crossover is
of considerable interest since experiments are now approaching the
Tonks regime
\cite{DePMcCWin99,BonBurDet01,GorVogLea01,DetHelRyy01,Greiner}. An
approximate theory in a strictly 1D model, the Lieb-Liniger (LL)
delta-function Bose gas \cite{LL} in a harmonic trap, has been
given recently by Dunjko, Lorent, and Olshanii \cite{DJM}, but it
is unclear how to extend their approach to 3D as it is rooted in
the 1D LL model. The condensed fraction $N_{0}/N$ is expected to
be small for narrow waveguides due to the above-described $N^p$
behavior of $N_0$. Variational approaches such as
Hartree-Fock-Bogoliubov (HFB) theory or its forerunner, the
variational pair theory or Girardeau-Arnowitt (GA) theory of
many-boson systems \cite{GA}, suggest themselves here. Here we
generalize the GA theory so as to treat the gradual onset of
``fermionization'' as the Tonks limit is approached and apply it
to the theory of the BEC-Tonks crossover. We also illustrate how
the theory can be extended to 3D by allowing for variable
transverse confinement.

{\em Number-conserving pair theory}: The most general Bose pairing
state involves excitation of pairs to arbitrary excited
(uncondensed) states $(n,m)$ by repeated  application of pair
excitation operators $\hat{a}_{n}^{\dagger}\hat{a}_{m}^{\dagger}
\hat{a}_{0}^2$ to the completely condensed $N$-particle state
$|N\rangle =(N!)^{-1/2}(\hat{a}_{0}^{\dagger})^{N}|0\rangle$. Here
we employ a number-conserving formulation \cite{GA,GC,Arnoproc} and define
{\em unitary} condensate annihilation and creation operators
${\hat{\beta}}_{0}={({\hat{N}}_{0}+1)}^{-1/2}\hat{a}_0$ and
${\hat{\beta}}_{0}^{\dagger}=\hat{a}_0^{\dagger}{({\hat{N}}_{0}+1)}^{-1/2}$
which commute with each other and with all annihilation and
creation operators $\hat{a}_{n}$ and $\hat{a}_{n}^{\dagger}$ for
noncondensed atoms ($n\ne 0$); here
$\hat{N}_{0}=\hat{a}_{0}^{\dagger}\hat{a}_{0}$. Then the
number-conserving generalized pair creation operator is
$\hat{a}_{n}^{\dagger}\hat{a}_{m}^{\dagger}{\hat{\beta}}_{0}^{2}$.
In the case of a sufficiently symmetrical trap the general pairing
state can be reduced to a $\pm k$ pairing form
\begin{equation}
|\Phi_0\rangle=\mbox{const.}e^{-\hat{F}}|N\rangle \ ,\
\hat{F}=\frac{1}{2}\sum_{k\neq 0}\Delta_{k}
\hat{a}_{k}^{\dagger}\hat{a}_{-k}^{\dagger}\hat{\beta}_{0}^{2} .
\end{equation}
by a suitable choice of orbitals $u_k$ where $u_{-k}$ denotes the time
reversal conjugate of $u_k$ and $\Delta_{k}$ is real and even.
This state is the
vacuum of number-conserving Bose and Fermi quasiparticle annihilation
operators \cite{GA,GC}:
\begin{equation}
\hat{\xi}_{k}=(1-\Delta_{k}^{2})^{-1/2}
(\hat{\beta}_{0}^{\dagger}\hat{a}_{k}
+\Delta_{k}\hat{a}_{-k}^{\dagger}\hat{\beta}_{0}) \ , \ k\neq 0 ,
\end{equation}
and the general pairing state can be written as
$|\Phi_0\rangle=\hat{U}|N\rangle$ where
$\hat{U}^{-1}\hat{\xi}_{k}\hat{U}=\hat{a}_{k}$.

{\em Toroidal and long cylindrical geometries}: A convenient
geometry for discussing the crossover is a toroidal trap of high
aspect ratio $R=L/\ell_{0}$ where $L$ is the toroid circumference
and $\ell_{0}$ the transverse oscillator length
$\ell_{0}=\sqrt{\hbar/m\omega_0}$ with $\omega_0$ the frequency of
transverse oscillations, assumed to be harmonic. The transverse
trap potential is assumed to be symmetric about an axis consisting
of a circle on which the trap potential is minimum. The
longitudinal (circumferential) motion can be described by a 1D
coordinate $x$ in terms of plane-wave orbitals
$\phi_{k}(x)=L^{-1/2}e^{ikx}$ satisfying periodic boundary
conditions with periodicity length $L$, with allowed longitudinal
quantum numbers $k_{j}=2\pi n_{j}/L$ with $n_{j}=0,\pm 1,\pm
2,\cdots$. The corresponding 3D orbitals are taken to be
$u_{k}({\bf r})=\phi_{k}(x)\phi_{tr}(\rho)$ where $\rho$ is a
transverse radial coordinate measured from the central circle of
the toroid; these are cylindrical coordinates with cylinder axis
bent into a circle of circumference $L$. This geometry can equally
well be interpreted as an infinitely long, straight cylindrical
waveguide with periodic boundary conditions in the longitudinal
direction. The creation operators in the variational trial state
$|\Phi_0\rangle$ refer to the 3D orbitals $u_{k}({\bf r})$. Use of
a single transverse orbital is justified both in the Tonks limit,
where transverse excitations are absent, and in the GP limit,
where BEC is almost complete and there is a single orbital
determined by the GP equation.

{\em Interatomic interaction}: We use the usual Fermi
pseudopotential $v({\bf r}_{ij})=4\pi a(\hbar^{2}/m)\delta({\bf
r}_{ij})$ and assume that the s-wave scattering length $a$ is
positive. This leads to a well-defined problem in 1D, the LL model
\cite{LL}. Our toroidal system is ``almost 1D'' since the
transverse dimensions are confined, and we find that the
variational problem with the Fermi pseudopotential does not
encounter the difficulties (divergences and poorly-posed
variational problem) \cite{Huang1,Huang2} found in the 3D case. A
gapless theory with more complicated pseudopotential, e.g.
\cite{OP}, is not warranted here since we are concerned with the
ground state, collective phonon excitations of the GA theory are
gapless \cite{Huang2,HM}, and the quasiparticle gap removes an
unphysical low-momentum divergence in the 1D Bogoliubov theory.

{\em Pair Hamiltonian}: The second-quantized
many-boson Hamiltonian with units $\hbar=m=1$ is
\begin{equation}
\hat{H}=\sum_{k}\epsilon_{k} \hat{a}_{k}^{\dagger}\hat{a}_{k}
+\frac{g}{2L}\sum_{qkk^{'}}\hat{a}_{k+q}^{\dagger}
\hat{a}_{k^{'}-q}^{\dagger}\hat{a}_{k^{'}}\hat{a}_k ,
\end{equation}
with single particle energy
$\epsilon_{k}=(k^{2}/2)+\epsilon_{tr}$, transverse mode
energy
\begin{equation}
\epsilon_{tr}=\int_{0}^{\infty}\phi_{tr}^{*}\left(-\frac{1}{2\rho}
\frac{\partial}{\partial\rho}\rho\frac{\partial}{\partial\rho}
+\frac{\omega_0^2}{2}\rho^{2}\right)\phi_{tr}2\pi\rho d\rho ,
\end{equation}
where the transverse orbital is normalized according to
$\int_{0}^{\infty}|\phi_{tr}(\rho)|^{2}2\pi\rho d\rho =1$, and
interaction matrix element $g=4\pi
a\int_{0}^{\infty}|\phi_{tr}(\rho)|^{4}2\pi\rho d\rho$. The only
interaction terms with nonzero expectation value in the state
$|\Phi_0\rangle$ are those expressible in terms of momentum
occupation number operators
$\hat{N}_{k}=\hat{a}_{k}^{\dagger}\hat{a}_{k}$ and pair operators
$\hat{a}_{-k}\hat{a}_{k}$ and
$\hat{a}_{k}^{\dagger}\hat{a}_{-k}^{\dagger}$, and of those the
transverse terms and the interaction terms with $q=0$ sum to
$\epsilon_{tr}\hat{N}+(g/2L)\hat{N}(\hat{N}-1)$ where
$\hat{N}=\sum_{k}\hat{N}_{k}$ is the total particle number
operator. Replacing this operator by its eigenvalue $N$, one
obtains a pair Hamiltonian \cite{GA}
\begin{eqnarray}
\hat{H}_{P}&=&\epsilon_{tr}N+(g/2L)N(N-1) \nonumber \\
&+&\sum_{k\ne 0}[(k^{2}/2) +(g/L)\hat{N}_{0}]\hat{N}_{k} \nonumber
\\ &+&\frac{g}{2L}\sum_{k\ne 0}
\{[\hat{N}_{0}(\hat{N}_{0}-1)]^{1/2}(\hat{\beta}_{0}^{\dagger})^{2}
\hat{a}_{-k}\hat{a}_{k}+\mbox{h.c.}\} \nonumber \\
&+&\frac{g}{2L}\sum_{kk^{'}\ne 0}[\hat{N}_{k}\hat{N}_{k^{'}}
\nonumber \\ &+&(1-\delta_{kk^{'}}-\delta_{k,-k^{'}})
\hat{a}_{k}^{\dagger}\hat{a}_{-k}^{\dagger}
\hat{a}_{-k^{'}}\hat{a}_{k^{'}}] ,
\end{eqnarray}
whose expectation value in the state $|\Phi_{0}\rangle$ is identical with
that of the full Hamiltonian (3).

{\em Bogoliubov theory}: In the case of a an untrapped 3D system
the Bogoliubov theory is valid at low densities. In the opposite
limit of a strictly 1D system with delta-function interaction (LL
model), the Bogoliubov theory reproduces the leading terms in the
exact ground state energy in the limit of {\em high} densities
\cite{LL}. However, the Bogoliubov theory is not fully consistent
either in the strictly 1D case or in our geometry: We find a
Bogoliubov quasiparticle energy
$\omega_{k}=|k|\sqrt{n_{0}g+k^{2}/4}$ and momentum distribution
function behaving like $|2k|^{-1}\sqrt{gn_{0}}$ at low momentum,
where $n_0$ is the mean value of $\hat{N}_{0}/L$. This leads to a
logarithmic divergence of the depletion integral (fractional
occupation of orbitals with $|k|>0$). We therefore proceed
directly to the variational theory.

{\em Variational pair (GA) theory}: The variational ground state
energy $E_{0P}$ is most easily evaluated by recalling that
$|\Phi_0\rangle$ is the vacuum of number-conserving quasiparticle
annihilation operators (2), applying Wick's theorem after
rewriting $\hat{H}_P$ in terms of quasiparticle annihilation and
creation operators via the inverse transformation
$\hat{\beta}_{0}^{\dagger}\hat{a}_{k}=(1-\Delta_{k}^{2})^{-1/2}
(\hat{\xi}_{k}-\Delta_{k}\hat{\xi}_{-k}^{\dagger})$. We pass to
the thermodynamic limit in the longitudinal direction, i.e.
$N\to\infty$, $L\to\infty$, $N/L=n$ (linear number density). Then
the operator square root in the third line of (5) may be replaced
by $\hat{N}_0$ with negligible error and $\hat{N}_0$ eliminated
from the second and third lines via the identity
$\hat{N}_0=N-\sum_{k\ne 0}\hat{N}_k$ valid for eigenstates of
total particle number with eigenvalue $N$. Applying Wick's
theorem, passing to the thermodynamic limit via $\sum_{k\ne
0}\to(L/2\pi)\int_{-\infty}^{\infty}dk$, and noting that all
integrands are even functions of $k$, one finds eventually
\begin{eqnarray}
e_{0P}&=&\epsilon_{tr}+\frac{ng}{2} +\frac{1}{\pi
n}\int_{0}^{\infty}[(\frac{k^2}{2}+n_{0}g +\frac{1}{2}I_{2})N_{k}
\nonumber\\
&-&(n_{0}g-\frac{1}{2}I_{1})\frac{\Delta_{k}}{1-\Delta_{k}^{2}}]dk
,
\end{eqnarray}
where $e_{0P}=E_{0P}/N$ is the energy per particle, $n_{0}=nf$ is
the condensate number density with condensate fraction
$f=1-\frac{1}{N}\sum_{k\ne 0}N_k = 1-\int_0^\infty \frac{N_k}{\pi
n}dk$, the momentum distribution function is
$N_{k}=\Delta_{k}^{2}/(1-\Delta_{k}^{2})$, and
\begin{equation}
I_{1}=\frac{g}{\pi}\int_{0}^{\infty}\frac{\Delta_{k}}{1-\Delta_{k}^{2}}dk
\quad , \quad
I_{2}=\frac{g}{\pi}\int_{0}^{\infty}N_{k}dk .
\end{equation}
$I_2=ng(1-f)$ and need not be evaluated separately.
These equations are in one-one correspondence with Eqs. (21) and (22) of
GA \cite{GA} via the correspondences $\Delta_{k}\rightarrow\phi(k)$,
$n_{0}g\rightarrow\rho_{0}\nu(k)$, $I_{1}\rightarrow I_{1}(k)$, and
$I_{2}\rightarrow I_{2}(k)$, but here the integrals $I_1$ and $I_2$ are
independent of $k$ due to our use of the Fermi pseudopotential.
Minimizing the
expression (6) by setting its functional derivative with respect to
$\Delta_{k}$ equal to zero, taking into account the dependence of $n_{0}$,
$I_1$, and $I_2$ on $\Delta_k$, one finds
\begin{equation}
(n_{0}g-I_{1})(1+\Delta_{k}^{2})-2[(k^{2}/2)+n_{0}g+I_{1}]\Delta_{k}=0
,
\end{equation}
whose solution for $\Delta_k$ is
\begin{equation}
\Delta_{k}=(n_{0}g-I_{1})^{-1}[(k^{2}/2)+n_{0}g+I_{1}-\omega_{k}]
,
\end{equation}
where $\omega_k$ is the quasiparticle energy
\begin{equation}
\omega_{k}=[k^{2}(n_{0}g+I_{1})+(k^{4}/4)+4n_{0}gI_{1}]^{1/2} .
\end{equation}
Substitution of this expression for $\Delta_k$ back into the
definitions of $I_{1}$ and $I_2$ leads to integrals which can be
evaluated in closed form
\begin{eqnarray}
I_{1}&=&\frac{g(n_{0}g-I_{1})}{2\pi}
\int_{0}^{\infty}\frac{dk}{\omega_{k}}=\frac{g(n_{0}g-I_{1})}{2\pi}
\frac{K}{\sqrt{n_{0}g}} ,\\ I_{2} &=& ng(1-f)=\frac{g}{2\pi}
\int_{0}^{\infty}\frac{dk}{\omega_{k}}\left
(\frac{k^{2}}{2}+n_{0}g+I_{1}-\omega_k \right )\nonumber\\ &=&
\frac{g}{2\pi}\sqrt{n_{0}g} [(1+\lambda)K-2E] , \label{I1I2}
\end{eqnarray}
where $\lambda=I_{1}/n_{0}g$, and $K=K(\sqrt{1-\lambda})$ and
$E=E(\sqrt{1-\lambda})$ are complete elliptic integrals \cite{GR}.
Here we have made use of the fact that $I_{1}>0$, since otherwise
$\omega_k$ would become imaginary at small $k$, which further
requires that $n_0g>I_1$. The ground state energy per particle is
finally found to be
\begin{equation}
e_{0P}=\epsilon_{tr}+\frac{ng}{2}-f(I_{1}-I_{2}-I_{3})
+\frac{(I_{1}^{2}+I_{2}^{2})}{2ng} ,
\end{equation}
where
\begin{eqnarray}
I_{3}&=&\frac{1}{4\pi n_{0}}
\int_{0}^{\infty}k^{2}\left[\frac{(k^{2}/2)+n_{0}g+I_{1}}{\omega_{k}}
-1\right]dk \nonumber\\
&=&\frac{\sqrt{g/n_{0}}}{3\pi}[(n_{0}g+I_{1})E -2I_{1}K]  .
\end{eqnarray}
\begin{figure}
\epsfxsize 3 in \epsfbox{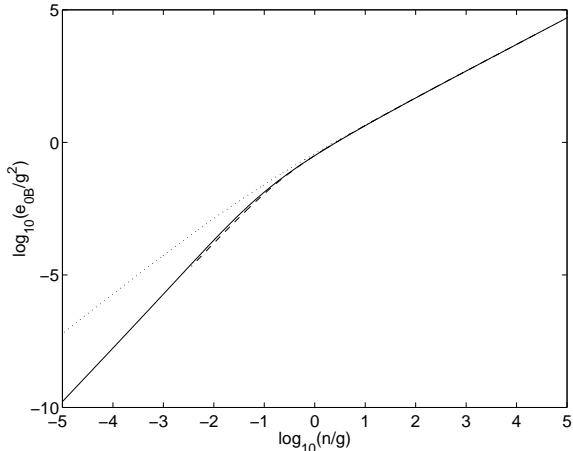} \caption{Log-Log
plot of the scaled energy per particle versus the scaled linear
number density $n/g$: Hybrid Bose-Fermi variational theory
$e_{0B}/g^2$ (solid line), exact LL solution $e_{LL}/g^2$ (long
dash line), $e_{0P}/g^2$ from the GA theory (short dash line).}
\label{Fig:one}
\end{figure}

The short dash line in Fig. 1 shows a log-log plot of the scaled
energy per particle $e_{0P}/g^2$ as a function of scaled linear
density $n/g$ calculated using the GA theory in Eqs. (6)-(14) for
constant coupling $g$. Fixing $g$ corresponds to the
regime where the transverse orbital is frozen as the lowest mode
of the harmonic trap, in which case our model coincides with that
of Lieb and Liniger \cite{LL}. The long dash line in Fig. 1,
almost indistinguishable from the solid line, is the
scaled energy per particle $e_{LL}(\gamma)/g^2=(n/g)^2e(\gamma)/2$
from the LL theory. There is excellent agreement between the GA
theory and LL theory in the high density regime $n/g>1$, where the
LL ground state energy is known to agree with Bogoliubov theory
\cite{Bog,LL}, but for low densities $n/g<1$ the predicted
energies diverge.

{\em Hybrid Bose-Fermi variational theory}: The failure of the GA
theory for $n/g<1$ can be traced to ``fermionization'' in which
bosonic atoms become impenetrable at low densities \cite{Olshanii}; 
this is the 1D Tonks gas limit where Bose-Fermi mapping \cite{map} applies.
A variational method capable of treating ``partial fermionization'' is 
suggested to treat the BEC-Tonks crossover. 
In the crossover region the system is intermediate between a BEC and
a fermionized Tonks gas, so we model it as an interpenetrating mixture
of a BEC of $wN$ bosons with pair theory energy functional $e_{0P}$
and $(1-w)N$ fermions with energy functional $e_{0F}$. Assuming additivity
of the Bose and Fermi energies as in the theory of ideal mixtures
\cite{tW}, one obtains an approximate energy functional (total energy
per particle)
\begin{equation}
e_{0B}(n)\approx we_{0P}(wn)+(1-w)e_{0F}((1-w)n)\ .
\end{equation}
The Fermi energy functional is just 
that of the ideal Fermi gas, $e_{0F}(n)=\pi^2n^2/6$, since a contact
interaction of spinless fermions is equivalent to no interaction.
Then for a fixed linear density $n$ one obtains 
$e_{0B}(n)\le we_{0P}(wn)+(1-w)^3\pi^2n^2/6$ and this can be numerically 
minimized with respect to $w$, using $e_{0P}$ from Eq. (13). The solid line 
in Fig. 1
shows the resultant scaled ground state energy $e_{0B}/g^2$ versus scaled
linear density $n/g$ for constant $g$, and excellent agreement is seen 
with the LL theory (long dash line) over the full range of densities. In
particular, for low densities $n/g$ the energy per particle
correctly approaches that of a free Fermi gas \cite{Olshanii,map,LL}.
One notes from Fig. 1 that the variational energy lies above the exact
LL energy, providing an {\it a posteriori} justification of the minimization
with respect to variation of $w$.

\begin{figure}
\epsfxsize 3 in \epsfbox{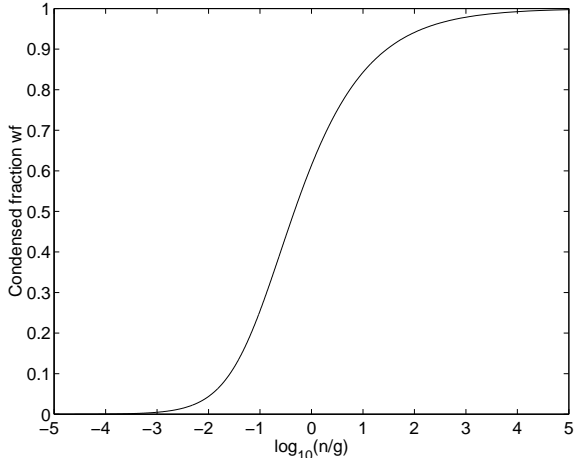} \caption{Condensed
fraction $wf$ versus the logarithm of the scaled linear number
density $log_{10}(n/g)$.} \label{Fig:two}
\end{figure}

{\em BEC-Tonks crossover:} Insofar as calculation of the energy via (15) is
concerned, the system behaves as if $(1-w)N$
interacting bosons have been replaced by free fermions. The fermionic
contribution dominates when $n/g\ll 1$ where the bosons are impenetrable
and the Bose-Fermi mapping theorem \cite{map} applies, whereas the GA
theory is accurate in the opposite limit $n/g>1$. In the crossover
region $n/g\sim 1$ we define an effective Bose condensed fraction by
$wf=N_{0}/N$ where $f=N_{0}/N_B$ and $N_{B}=wN$.
Figure \ref{Fig:two} shows this
condensed fraction as a function of the the scaled density
for constant $g$, and a crossover occurs at $n^*/g=2.17$ where
$wf=0.5$: This crossover condition $n^*/g=2.17$ is very close to
Olshanii's prediction $N^*=L/\pi|a_{1D}|$ for the maximum number
of atoms to form a 1D Tonks gas, where we identify $N^*/L=n^*$ and
$g=2/\pi|a_{1D}|, a_{1D}$ being the 1D scattering length
\cite{Olshanii}. Petrov et al. derived the same condition for the
crossover from a Tonks gas to a 1D quasicondensate \cite{Petrov}.

For an ideal Tonks gas $w\rightarrow 0$ which requires
$n/g\rightarrow 0$, but as a measure of the density needed to
approach a Tonks gas if we require 90\% or more of the atoms to be
fermionized (w=0.1) we need to satisfy $n/g<0.026$. By comparison,
for a condensed fraction greater than 90\% $(wf>0.9)$ we require
$n/g>30$. By comparison, using hydrostatic equations applied to a
trapped LL model \cite{DJM},  Dunjko {\it et al.} find that the
density profile of the trapped gas is close to the Thomas-Fermi
solution for $n/g=14$ $(\eta=9)$ for which $wf=0.85$, and close to
that for a Tonks gas for $n/g\approx 0.1$ $(\eta=0.07)$ for which
$wf=0.25$, where $\eta=n|a_{1D}|$ in their notation. There is
therefore consistency between our results even though we do not
consider a gas with longitudinal trapping.

In summary, we have developed a hybrid Bose-Fermi variational
theory that accurately describes the BEC-Tonks crossover in 1D. A
key virtue of our approach is that it may be extended to 3D: Here
we have assumed tight confinement and taken $g=4\pi
a\int_{0}^{\infty}|\phi_{tr}(\rho)|^{4}2\pi\rho d\rho$ as a
constant whose value for the unperturbed transverse ground orbital
is $g=2a/\ell_{0}^2$ where $\ell_{0}=1/\sqrt{\omega_0}$
\cite{Petrov}. More generally, minimization of the ground state
energy by variation of $\phi_{tr}$ subject to the normalization
constraint leads to the following generalized GP equation:
\begin{eqnarray}
\mu\phi_{tr} &=& -\frac{1}{2}\left
(\frac{\partial^2}{\partial\rho^2} +
\frac{1}{\rho}\frac{\partial}{\partial\rho} \right )\phi_{tr} +
\frac{1}{2}\omega_0^2\rho^2\phi_{tr} \nonumber \\ &+& 4\pi
an(2-f^2-2\lambda f+f^2\lambda^2)|\phi_{tr}|^2\phi_{tr} ,
\end{eqnarray}
where $\mu$ is the chemical potential. Solving this GP equation
for $\phi_{tr}$ to obtain $g$, and then solving self-consistently
with Eqs. (6)-(14) will allow for the study of the crossover from
1D to 3D and will be the subject of future work. \vspace{0.2cm}

\noindent We thank Prof. Maxim Olshanii for useful discussions, 
and Vanja Dunjko for making available, on the Web,  the results of 
his evaluation of the energy functional of \cite{LL}; see 
ref. 19 of \cite{DJM}. This work was supported by Office 
of Naval Research grant N00014-99-1-0806 and by the US 
Army Research office.


\begin{references}
%
\bibitem{Gross}         E.P. Gross, \Journal{\NC}{20}{454}{1961};
E.P. Gross, \Journal{\JMP}{4}{195}{1963}.
%
\bibitem{Pitaevskii}    L.P. Pitaevskii, \Journal{\ZETF}{40}{646}{1961}
[\Journal{\JETP}{13}{451}{1961}].
%
\bibitem{Bog}           N.N. Bogoliubov, \Journal{\JPUSSR}{11}{23}{1947}.
%
\bibitem{DGPS}          F. Dalfovo, S. Giorgini, L.P. Pitaevskii, and S. Stringari,
\Journal{\RMP}{71}{463}{1999}, particularly Sec. IV.A, pp. 480 ff.
%
\bibitem{Olshanii}      M. Olshanii, \Journal{\PRL}{81}{938}{1998}.
%
\bibitem{Petrov}        D.S. Petrov {\it et al.}, \Journal{\PRL}{85}{3745}{2000}.
%
\bibitem{map}           M. Girardeau, \Journal{\JMP}{1}{516}{1960}; \Journal{\PR}
                        {139}{B500}{1965}.
%
\bibitem{GW3}           M.D. Girardeau, E.M. Wright, and J.M. Triscari,
                        \Journal{\PRA}{63}{033601}{2001}.
%
\bibitem{DePMcCWin99}   M. T. Depue {\it et al.}, \Journal{\PRL}{82}{2262}{1999}.
%
\bibitem{BonBurDet01}   K. Bongs {\it et al.}, \Journal{\PRA}{63}{031602}{2001}.
%
\bibitem{GorVogLea01}   A. Gorlitz {\it et al}, cond-mat/0104549 (2001).
%
\bibitem{DetHelRyy01}   S. Dettmer {\it et al}, cond-mat/0105525 (2001).
%
\bibitem{Greiner}       M. Greiner {\it et al.}, cond-mat/0105105 (2001).
%
\bibitem{LL}            E.H. Lieb and W. Liniger, \Journal{\PR}{130}{1605}{1963}.
%
\bibitem{DJM}           V. Dunjko, V. Lorent, and M. Olshanii, \Journal{\PRL}{86}{5413}{2001}.
%
\bibitem{GA}            M. Girardeau and R. Arnowitt, \Journal{\PR}{113}{755}{1959}.
%
\bibitem{GC}            M.D. Girardeau, \Journal{\PRA}{58}{775}{1998}.
%
\bibitem{Arnoproc}M.D. Girardeau, in {\it Relativity, Particle Physics, and
Cosmology
(Proceedings of the Richard Arnowitt Fest, Texas A\&M University,
5-7 April 1998)}, ed. Roland E. Allen (World Scientific, 1999), pp. 190 ff.
%
\bibitem{Huang1}        K. Huang, Int. J. Mod. Phys. (Singapore) {\bf A4}, 1037 (1989).
%
\bibitem{Huang2}        K. Huang and P. Tommasini, J. Res. Nat. Inst. Standards
                        and Tech. {\bf 101}, 435 (1996).
%
\bibitem{OP}            M. Olshanii and L. Pricoupenko, cond-mat/0101275 (2001).
%
\bibitem{HM}            P.C. Hohenberg and P.C. Martin, \Journal{\APNY}{34}{291}{1965}.
%
\bibitem{LHY}           T.D. Lee, K. Huang, and C.N. Yang, \Journal{\PR}{106}{1135}{1957}.
%
\bibitem{GR}            I.S. Gradshteyn and I.M. Ryzhik, {\it Table of Integrals,
                        Series, and Products} (Academic Press, New York,
                        1980).
%
\bibitem{tW}D. ter Haar and H.N.S. Wergeland, {\it Elements of Thermodynamics}
(Addison-Wesley Publishing Co., Reading, Mass., 1966), pp. 85 ff.
%
\end{references}
\end{document}